\documentclass[aps,pre,epsf,superscriptaddress,amsmath,amssymb,amsfonts,twocolumn,showpacs]{revtex4-1}

\usepackage{graphicx}
\usepackage{epsfig}
\usepackage{dcolumn}
\usepackage{bm}
\usepackage{braket}
\usepackage{amsmath}
\usepackage{color} 

\begin{document}

\title{Dark-bright soliton pairs: bifurcations and collisions}
 
\author{G. C. Katsimiga}
\affiliation{Zentrum f\"{u}r Optische Quantentechnologien,
Universit\"{a}t Hamburg, Luruper Chaussee 149, 22761 Hamburg,
Germany}  

\author{P. G. Kevrekidis}
\affiliation{Department of Mathematics and Statistics, University
of Massachusetts Amherst, Amherst, MA 01003-4515, USA}

\author{B. Prinari}
\affiliation{Department of Mathematics, University of Colorado Colorado Springs, 
Colorado Springs, Colorado 80918, USA}

\author{G. Biondini}
\affiliation{Department of Mathematics and Department of Physics, State University 
of New York, Buffalo, New York 14260, USA}

\author{P. Schmelcher}
\affiliation{Zentrum f\"{u}r Optische Quantentechnologien,
Universit\"{a}t Hamburg, Luruper Chaussee 149, 22761 Hamburg,
Germany} \affiliation{The Hamburg Centre for Ultrafast Imaging,
Universit\"{a}t Hamburg, Luruper Chaussee 149, 22761 Hamburg,
Germany}

\date{\today}

\begin{abstract}
The statics, stability and dynamical properties of
dark-bright soliton pairs are investigated motivated by applications in a
homogeneous system of two-component 
repulsively interacting Bose-Einstein condensate.
One of the intra-species interaction coefficients is used as the relevant 
parameter controlling the deviation from the integrable Manakov limit.
Two different families of stationary states are identified consisting of dark-bright
solitons that are either antisymmetric (out-of-phase) 
or asymmetric (mass imbalanced) with respect to their bright soliton. 
Both of the above dark-bright configurations coexist at the integrable limit of equal intra- and 
inter-species repulsions and are degenerate in that limit. However,
they are found to bifurcate from it in a transcritical bifurcation.  
The latter interchanges the stability properties of the bound dark-bright
pairs rendering the antisymmetric states 
unstable and the asymmetric ones stable past the associated critical point
(and vice versa before it). 
Finally, on the dynamical side, it is found that large kinetic energies and 
thus rapid soliton collisions are essentially unaffected by the intra-species variation, while cases involving 
near equilibrium states or breathing dynamics are significantly modified
under such a variation.
\end{abstract}

\maketitle

\section{Introduction}

Multi-component Bose-Einstein condensates (BECs) and the nonlinear excitations 
that arise in them have been a focal research point over
the past two decades since their experimental realization~\cite{gpes1,gpes2}.
Among these excitations, dark-bright (DB) solitons constitute a fundamental
example~\cite{frantzkevre},
whose experimental realization in a $^{87}$Rb mixture~\cite{hamburg} 
has triggered a new era of investigations regarding the stability
and interactions of these matter waves both with each other~\cite{yan}, 
as well as induced by the external traps~\cite{ba,hamner,middelkamp,alvarez}.

Within mean-field theory the static and dynamical properties of such states are 
well described by a system of coupled Gross-Pitaevskii equations~\cite{gpes1,gpes2}. 
The latter is a variant of the so-called defocusing (repulsive) vector nonlinear     
Schr{\"o}dinger equation~\cite{nls,siambook}, to which it reduces in the absence of a confining potential.
In this homogeneous setting, single DB solitons exist as exact analytical 
solutions when the repulsive interactions within (intra-) and between (inter-) 
the species are of equal strength; this is the integrable,
so-called Manakov limit~\cite{manakov}. 
In this setting, multiple solitonic states, both static and travelling ones,
have been analytically derived 
by using the inverse scattering transform (IST) considering both trivial~\cite{manakov}, or more recently, 
non-trivial boundary conditions~\cite{prinari,biondini} allowing also for
energy (and phase) exchanges 
between the bright soliton components. Additionally, the
Hirota method has been used to explore
different families of DB soliton solutions ranging from perfectly antisymmetric (out-of-phase) to fully asymmetric
ones (mass imbalanced), with respect to their bright soliton
counterpart~\cite{shepkiv} (see also for a small sample
among numerous additional
studies the works~\cite{vdbysk1,ralak,parkshin,rajendran}).

As a matter of fact, given their versatility, BECs offer additional
layers of tunability, enabling the controllable departure
from this integrable Manakov limit.
In particular, exploiting the tunablility of both the inter- 
and intra-species scattering lengths that can be achieved in 
current experimental settings with the aid of
Feshbach resonances~\cite{inouye,roberts,donley,thalhammer,chin}, 
a new avenue opened towards exploring 
DB soliton interactions under parametric
variations~\cite{ef,et,ljpp,carr,ljpp1}.
This allowed to address the robustness of these matter waves 
in BEC mixtures with genuinely different scattering lengths.   
However, typically in realistic settings all three coefficients
of inter- and intra- component interactions are slightly different,
and hence it is of particular interest to explore how things
deviate from the integrable limit.

In the present work, our aim is to bring to bear the enhanced
understanding of the integrable limit that exists through the
recent works of~\cite{prinari,biondini} in order to controllably
appreciate how statics, bifurcations and dynamics are affected
upon deviations from this limit. More specifically, we will examine
stationary states at the integrable limit and how they (and their
respective stability properties) are modified upon deviation from
integrability. In the process, we will uncover an unusual example
of a transcritical bifurcation with symmetry involving bound
DB soliton pairs of two kinds: antisymmetric and asymmetric ones.
In the former, the bright components are out-of-phase whereas
in the latter they are imbalanced in terms of their respective
masses. We will then turn to dynamical states involving (from
the integrable limit) solutions with different speeds. We will
initialize such states in regimes close to and far from integrability
to observe the implications of non-integrability on them.
Our main conclusion there is that for states of high kinetic energy
(where the latter dominates the DB interaction) implications of 
the non-integrability are rather limited. However, for states of proximal
DBs with prolonged (or recurrent) interactions, non-integrability
can have a significant impact in the outcome of their collisions,
as we illustrate via suitable numerical computations.

The paper is organized as follows. In Sec. II we provide the setup 
of the multi-component system under consideration. In Sec. III
the static properties of two-DB soliton solutions upon varying the intra-species 
interactions are exposed. Sec. IV is devoted to studying the dynamical properties 
of these matter waves, while Sec. V contains our conclusions and future perspectives. 

\section{Model setup}
As our prototypical playground, we consider the following one-dimensional (1D)
system of coupled nonlinear Schr{\"o}dinger equations: 
\begin{eqnarray}
i \partial_t \psi_d  =& -&\frac{1}{2} \partial_{x}^2\psi_d  
+(|\psi_d|^2 + g_{12}|\psi_b|^2 -\mu_d) \psi_d,\nonumber \\
\label{nls1}
\\
i \partial_t \psi_b  =& -&\frac{1}{2} \partial_{x}^2\psi_b  
+ (g_{12}|\psi_d|^2 + g_{22}|\psi_b|^2- \mu_b) \psi_b.\nonumber \\
\label{nls2}
\end{eqnarray}
In the above equations, $\psi_d$ ($\psi_b$) is the wavefunction of the dark (bright)
soliton component while $\mu_d$ ($\mu_b$) is the corresponding chemical potential.
Furthermore, $g_{12}\equiv g_{12}/g_{11}$,
and $g_{22}\equiv g_{22}/g_{11}$ denote the rescaled interaction coefficients 
which are left to arbitrarily vary, spanning both the 
miscible and the immiscible regime of interactions.
Note that in this setting the
miscibility of the two components occurs when
$g_{12} \leqslant \sqrt{g_{22}}$, and refers to the absence  of 
phase separation between the species~\cite{aochui}.  
Additionally, Eqs.~(\ref{nls1})-(\ref{nls2}) stem from the corresponding BEC system 
assuming a setting without a longitudinal trap, but only with a transverse trap
of strength $\omega_{\perp}$. The coupling constants in 1D are
$g_{jk}=2\hbar\omega_{\perp} a_{jk}$, 
where $a_{jk}$ denote the $s$-wave scattering lengths (with
$a_{12}=a_{21}$) that account for collisions between atoms of 
the same ($j=k$) or different ($j \ne k$) species.
The aforementioned dimensionless 1D system also assumes the measuring of
densities {$|\psi_j|^2$}, 
length, time and energy in
units of $2a_{11}$, $a_{\perp} = \sqrt{\hbar/\left(m \omega_{\perp}\right)}$, 
$\omega_{\perp}^{-1}$ and $\hbar\omega_{\perp}$, respectively. 
We note that in the following all our results are presented in dimensionless units.

From here, $\mu_d$ can be also scaled out via the transformations: $t \rightarrow 
\mu_d t$, $x \rightarrow {\sqrt{\mu_d}}x$, $|\psi_{d,b}|^2 \rightarrow \mu_{d}^{-1} |u_{d,b}|^2$, and thus the system of 
equations (\ref{nls1})-(\ref{nls2}) acquires the following form 
\begin{eqnarray}
i \partial_t u_d + \frac{1}{2} \partial_{x}^2u_d  
-(|u_d|^2 + g_{12}|u_b|^2 -1) u_d &=& 0,
\label{deq11}
\\
i \partial_t u_b +\frac{1}{2} \partial_{x}^2u_b  
-(g_{12}|u_d|^2 + g_{22}|u_b|^2- \mu) u_b &=& 0,
\label{deq21}
\end{eqnarray}
where $\mu\equiv\mu_b/\mu_d$ is the rescaled chemical potential.
The above system of equations conserves the total energy 
\begin{eqnarray}
E &=& \frac{1}{2}\int_{-\infty}^{+\infty} dx  
\Big[ |\partial_{x} u_d|^2+|\partial_{x} u_b|^2+(|u_d|^2-1)^2 \Big. \nonumber \\
&+& \Big. g_{22}|u_b|^4 - 2\mu |u_b|^2 + 2g_{12} |u_d|^2 |u_b|^2 \Big],
\label{energy} 
\end{eqnarray}
as well as the total number of atoms 
\begin{eqnarray}
N&\equiv &N_d+N_b=\sum_{i=d,b}\int^{\infty}_{-\infty} dx|u_i|^2,
\label{ntot}
\end{eqnarray}
with $N_d$, $N_b$, denoting the number of atoms in the first and second component of the  
system of Eqs.~(\ref{deq11})-(\ref{deq21}) respectively.
$N_d$ and $N_b$ are also individually conserved.

\begin{figure*}[ht]
\centering
\includegraphics[scale=0.39]{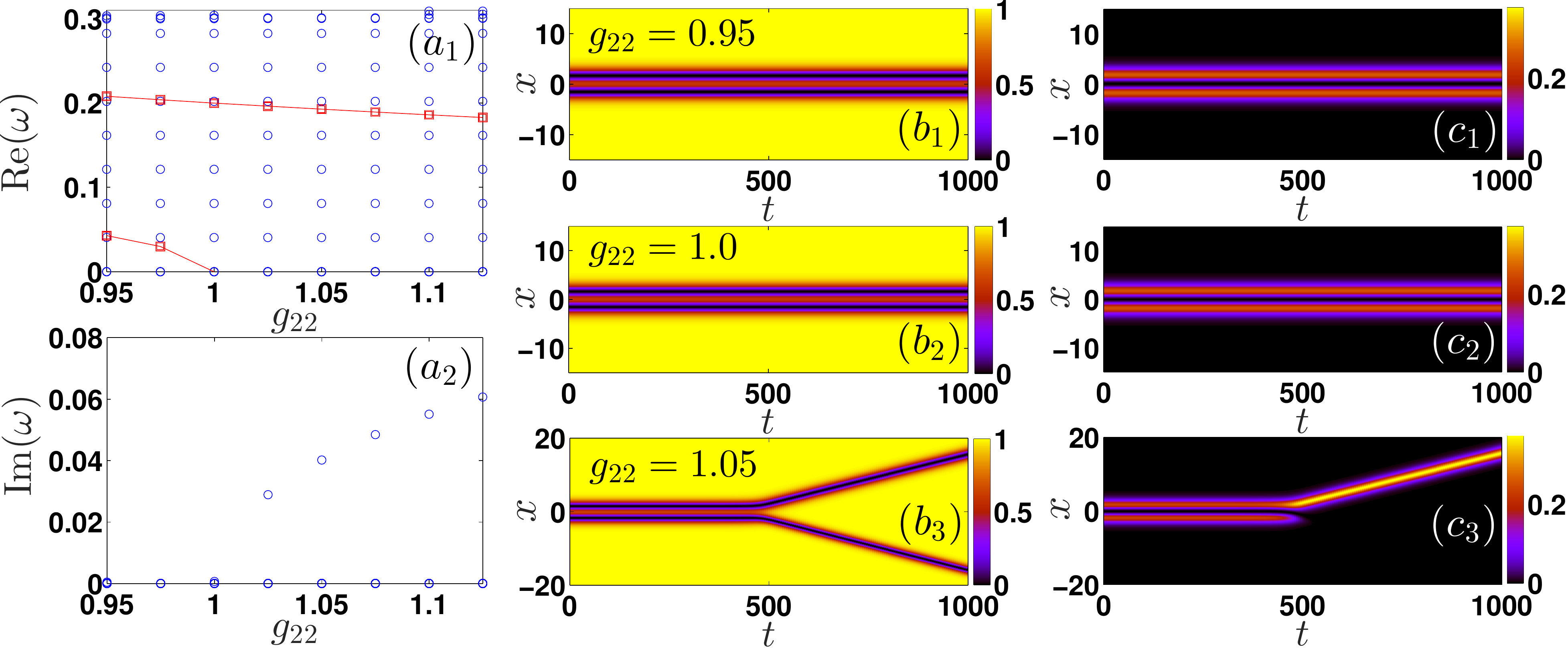}
\caption{(Color online): BdG (linearization) spectrum of stationary antisymmetric two-DB soliton states,  
upon varying $g_{22}$.  
Both $(a_1)$ the real, $\rm{Re}(\omega)$, and $(a_2)$ the imaginary part, $\rm{Im}(\omega)$,  
of the eigenfrequencies $\omega$ are shown as a function of the intra-species repulsion $g_{22}$. 
Upon increasing $g_{22}$ there exists a critical value, $g_{22_{cr}}=1$, above which the antisymmetric 
branch destabilizes. Note that since $g_{12}=1$, the critical point is the integrable limit, 
while the trajectories of the two anomalous modes (see text) appearing in the spectrum 
are shown with red squares (solid red lines are used as a guide to the eye). 
$(b_1)-(c_3)$ Spatio-temporal evolution of antisymmetric stationary two-DB states upon increasing $g_{22}$, 
verifying the BdG results. $(b_1)-(b_3)$ [$(c_1)-(c_3)$] correspond to the density $|u_d|^2$ [$|u_b|^2$] of the dark [bright] 
component. All quantities are expressed in dimensionless units.}   
\label{fig1}
\end{figure*}   

\section{Stability analysis of bound antisymmetric and asymmetric dark-bright pairs}
By considering the time-independent version of the aforementioned 
system of Eqs.~(\ref{deq11})-(\ref{deq21}), namely:
\begin{eqnarray}
u_d&=& -\frac{1}{2} \partial_{x}^2u_d  
+(|u_d|^2 + g_{12}|u_b|^2) u_d ,
\label{tidd}
\\
\mu u_b &=&-\frac{1}{2} \partial_{x}^2u_b  
+(g_{12}|u_d|^2 + g_{22}|u_b|^2) u_b,
\label{tidb}
\end{eqnarray}
bound states consisting of two DB solitons can be found in~\cite{ljpp,ljpp1} 
for out-of-phase or antisymmetric bright 
solitons for arbitrary nonlinear coefficients.

First, let us briefly recall what is known about the Manakov model.
In such a case the system possesses exact two-DB soliton solutions that can be obtained by using either Hirota's method studied in 
Ref.~\cite{shepkiv} or the more recent exact expressions  
found in Ref.~\cite{prinari} by using the inverse scattering transform (IST) but with non-trivial boundary conditions. 
In particular,  the exact static two-DB solutions can be
written in the following form:
\begin{eqnarray*}
 u_d &=&\frac{(1-a)\cosh(\xi_1+\xi_2)-(1+a)\cosh(\xi_1-\xi_2)}{(1-a)\cosh(\xi_1+\xi_2)+(1+a)\cosh(\xi_1-\xi_2)}, \\
 u_b &=&\frac{2(1-a^2) \sinh \xi_1}{(1-a)\cosh(\xi_1+\xi_2)+(1+a)\cosh(\xi_1-\xi_2)},
\end{eqnarray*}
where $\xi_1 = x - \delta_1$, $\xi_2 = a(x-\delta_2)$ and $\mu = 1- a^2/2$. 

There are three free parameters here: $a$, $\delta_1$ and $\delta_2$. One of them is due to the translational invariance:
Shifting $\delta_1 \rightarrow \delta_1 + \bar\delta$, $\delta_2 \rightarrow \delta_2 + \bar\delta$ only displaces the overall 
solution by $\bar \delta$,
so fixing the overall location of the pair one can fix $\delta_1 = -\delta_2 =: \delta$.
Then one is still left with a nontrivial two-parameter family, with parameters $a$ and $\delta$.
Fixing the chemical potential ratio $\mu$ fixes $a$ and vice versa (note that there is no loss of generality in assuming $a>0$ 
since $a \rightarrow -a$ only flips the global sign of $u_d$).
The parameter $a$ must satisfy $0 \leq a \leq 1$ and  accordingly $1/2 \leq \mu 
\leq 1$.
The second parameter $\delta$ can also be taken positive since $\delta \rightarrow - \delta$, $x\rightarrow -x$ leaves $u_d$, 
$u_b$ invariant (up to a global sign), 
so changing the sign of $\delta$ just performs a reflection about $x=0$, exchanging the two DBs. For $\delta=0$, $u_d( x)= 
u_d(-x)$, $u_b(x)=-u_b(-x)$, so the DB pair is antisymmetric.
As $\delta$ becomes larger, the asymmetry increases towards a dark/DB state and the equilibrium distance between the 
solitons increases.
In our simulations $\mu$ is fixed, so $a$ is fixed. 
But still, in the integrable limit there will be a one-parameter family of 
two-DB solutions of varying $\delta$, continuously ranging from 
perfectly antisymmetric to extremely asymmetric.

As we depart from integrability, such an explicit
expression is no longer available.
In light of that, the corresponding
stationary states were numerically obtained by means of a Newton's fixed
point iteration method 
and using as an initial guess for identifying the
numerically exact (up to a prescribed tolerance)
two-DB spatial profile  the following ansatz:
\begin{eqnarray}
\!\!\!\!\!\!
u_d(x)&=&\tanh\left[D(x-x_0)\right] \times \tanh\left[D(x+x_0)\right],
\label{dark}
\\
\!\!\!\!\!\!
u_b(x)&=&\eta {\rm sech}\left[D(x-x_0)\right] 
+\eta {\rm sech}\left[D(x+x_0)\right] e^{i\Delta \theta}. \nonumber\\
\label{bright}
\end{eqnarray}
In the above expressions $2x_0$ is the relative distance between the two DB solitons,
$D$ denotes their common inverse width, 
while $\Delta \theta$ is their relative 
phase within the bright component,
with $\Delta \theta=\pi$ ($\Delta \theta=0$) corresponding to out-of-phase (in-phase) bright solitons.
Note also that the (background) amplitude of the dark soliton component is unity, 
while $\eta$ denotes the amplitude of the bright soliton 
counterpart. 
Here, we will solely focus on the out-of-phase or antisymmetric case 
(as the in-phase case does not produce a bound state pair in the
homogeneous setting~\cite{ljpp}). 
Upon varying $g_{12}$ typically within the interval $0.75\leqslant g_{12}\leqslant 1.5$, 
while keeping both $\mu=2/3$ and $g_{22}=0.95$ fixed, 
earlier numerical studies~\cite{ljpp,ljpp1} 
showcased that antisymmetric 
two-DB states exist as stable configurations only within a bounded interval of the inter-species repulsion coefficient 
$g_{12}$ limited by critical points both in the miscible and in the immiscible regime, associated with a supercritical and a 
subcritical pitchfork bifurcation respectively. Furthermore, new families of solutions 
consisting of mass imbalanced DB pairs, i.e. different amplitudes between the bright soliton constituents, 
were found to bifurcate through pitchfork bifurcations
from the above obtained antisymmetric states. 

However, by fixing $g_{22} \neq 1$ in these previous works
it has not been possible to systematically approach the
Manakov limit of $g_{12}=g_{22}=1$. 
To address this important special limit, and explore the
effect of breaking the integrability, 
below we fix $g_{12}=1$ and perform a continuation in $g_{22}$, 
i.e., starting from $g_{22}=0.95$ (immiscible regime) 
up to $g_{22}=1$ and beyond, towards the miscible domain of interactions, 
considering the fate of both the antisymmetric and asymmetric 
bound states.
Note also that for the numerical findings to be presented below the rescaled chemical potential is fixed to $\mu=0.7$ 
in the dimensionless units adopted herein.    
In Figs.~\ref{fig1} $(a_1)-(a_2)$ 
the linearization, or so-called Bogolyubov-de Gennes (BdG), spectrum of the antisymmetric bound pairs is shown as a 
function of the nonlinear coefficient $g_{22}$.
This is obtained by expanding around an equilibrium configuration as
follows:
\begin{figure*}[ht]	
\includegraphics[scale=0.39]{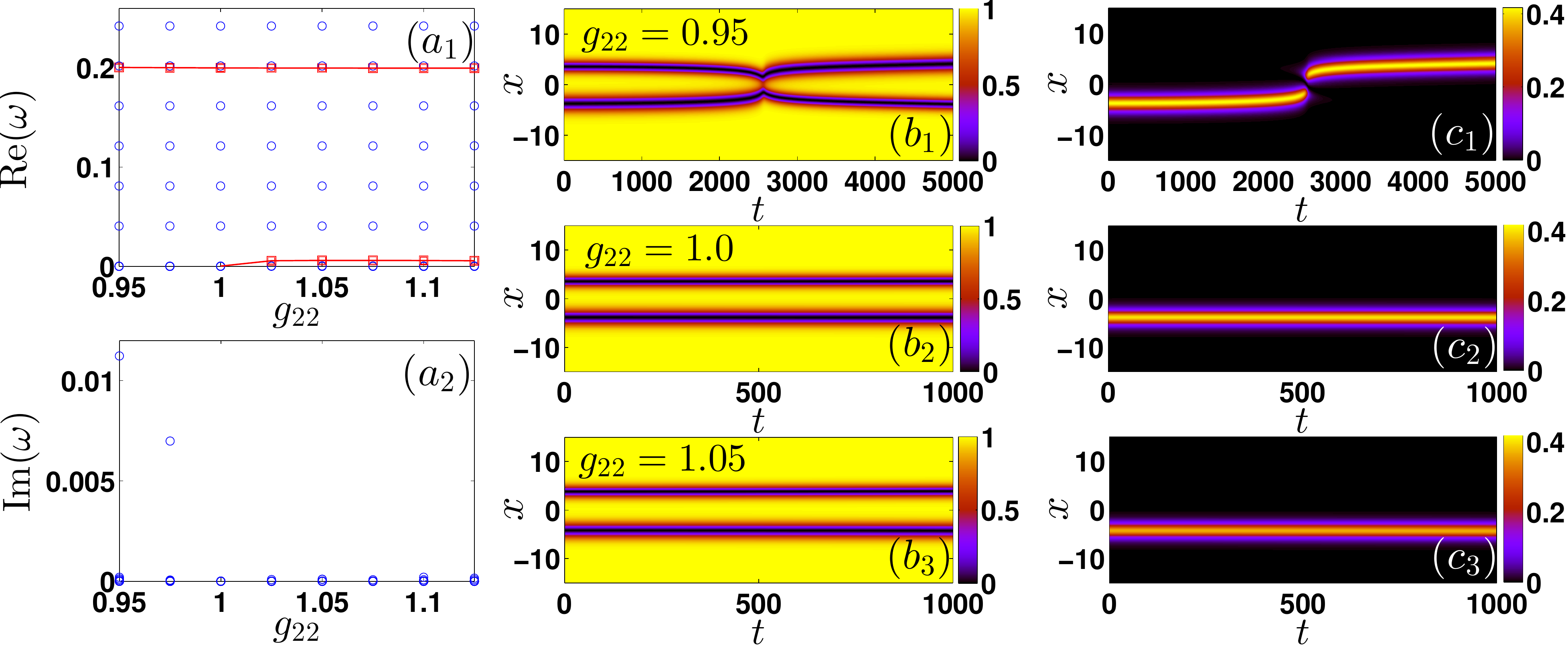}
\caption{(Color online): $(a_1)-(b_2)$ Same as in Fig.~\ref{fig1} but for the asymmetric two-DB states. 
Notice that in contrast to the antisymmetric states, these bound pairs are unstable for $g_{22}<1$, as can be seen from 
the non-zero imaginary part, $\rm{Im}(\omega)$, depicted in panel $(a_2)$. As the integrable limit is reached the 
bound pair stabilizes and remains robust even upon further increasing the nonlinear coefficient $g_{22}$. 
$(b_1)-(c_3)$ Space-time evolution of both the unstable ($g_{22}< 1$) and stable ($g_{22}\geqslant 1$) 
asymmetric stationary two-DB states upon increasing $g_{22}$. 
Panels $(b_1)-(b_3)$ 
[$(c_1)-(c_3)$] correspond to the density $|u_d|^2$ [$|u_b|^2$] of the dark [bright] component. 
All quantities are expressed in dimensionless units.}
\label{fig2}
\end{figure*}
%
\begin{eqnarray}
  u_d &=& u_d^{(eq)} + \left(a(x) e^{-i \omega t} + b^{\star}(x) e^{i \omega^{\star} t}
  \right),
  \label{bdg1}
  \\
 u_b &=& u_b^{(eq)} + \left(c(x) e^{-i \omega t} + d^{\star}(x) e^{i \omega^{\star} t}
  \right), 
  \label{bdg2}
\end{eqnarray}
where ``$\star$" stands for the complex conjugate.
Then the system for the eigenfrequencies $\omega$
(or equivalently eigenvalues $\lambda=i \omega$) and
eigenfunctions $(a,b,c,d)^T$ is numerically solved. If modes with purely
real eigenvalues (imaginary eigenfrequencies) or complex
eigenvalues (and thus eigenfrequencies) are identified, the configuration
is characterized as dynamically unstable. Moreover, there is a class
of modes that bears the potential to lead to instabilities. These are
the modes with negative so-called energy or Krein signature~\cite{krein}. The
relevant quantity is defined as
\begin{eqnarray}
  K= \omega \int  \Big(|a|^2 - |b|^2 + |c|^2 - |d|^2\Big) dx,
  \label{krein}
\end{eqnarray}
and can be directly evaluated on the basis of the eigenvector
$(a,b,c,d)^T$ and eigenfrequency $\omega$.
Both the real, $\rm{Re}(\omega)$, and the imaginary, $\rm{Im}(\omega)$, parts of the eigenfrequencies 
$\omega$ are depicted in Figs.~\ref{fig1} $(a_1)$ and $(a_2)$ respectively. 
Notice that in close contact with our previous findings~\cite{ljpp,ljpp1}, 
two anomalous (namely, bearing negative Krein signature) modes appear in the
linearization spectra and their trajectories 
are denoted with red squares [see Fig.~\ref{fig1} $(a_1)$]. 
Among these modes, the higher-lying one is found to be related 
to the out-of-phase vibration of the bound DB pair~\cite{ljpp}.
More importantly, the lower-lying anomalous mode is associated,
through its eigenvector, with a symmetry breaking in the bright
soliton component, resulting in mass imbalanced 
(with respect to their bright soliton counterpart) DB pairs
that we will trace later on in the dynamics. 
In both cases the aforementioned findings can be identified by adding the corresponding eigenvector
to the relevant antisymmetric DB solution. 
The background
(continuous, in the limit of infinite domain) spectrum is also
depicted in the same figure with blue circles. 
As it is observed, upon increasing $g_{22}$ towards the integrable limit the
frequencies of both of the anomalous modes 
decrease.
Following the lower-lying mode it becomes apparent that there exists a critical point $g_{22_{cr}}=1$ where this mode goes 
from linearly stable (for $g_{22}< g_{22_{cr}}$) to linearly unstable (for $g_{22}> g_{22_{cr}}$).  
Notice that exactly at the integrable limit this anomalous mode becomes neutrally stable,
while past $g_{22_{cr}}=1$  
it destabilizes as it is evident from the non-zero imaginary part presented in Fig.~\ref{fig1} $(a_2)$.

To verify the stability analysis results presented above,
the spatio-temporal evolution of both the stable and unstable antisymmetric DB pairs is computed and 
shown in Figs.~\ref{fig1} $(b_1)-(c_3)$. 
Notice that the antisymmetric configuration is only
slightly modified as $g_{22}$ is increased over the interval considered.
Minor differences, mostly in the amplitude of the bound pairs upon increasing $g_{22}$, can be inferred 
by inspecting the overall decrease of the norm of the bright component (see Fig.~3 below).
Here, Figs.~\ref{fig1} $(b_1)-(b_3)$ [$(c_1)-(c_3)$] correspond to the density of the dark [bright] soliton 
component. 
In particular, in all cases we use as an initial condition the numerically obtained stationary states
at selected values of $g_{22}$, i.e., below, at and above the associated critical point, 
and we numerically evolve the system 
of Eqs.~(\ref{deq11})-(\ref{deq21}) using a fourth order Runge-Kutta integrator. 
As anticipated from the aforementioned BdG outcome, for $g_{22}>g_{22_{cr}}=1$ the instability 
dynamically manifests itself via 
a dramatic mass redistribution between the bright 
soliton counterparts. The latter leads in turn to 
the splitting of the bound pair and results in asymmetric states with a dark and a DB soliton pair 
repelling  one another and moving towards the boundaries (cf. also with the
corresponding non-integrable cases in~\cite{ljpp,ljpp1}).

We now explore the same diagnostics for the asymmetric DB bound pairs that
are degenerate with the antisymmetric ones in the integrable limit.
These stationary asymmetric states are once again numerically identified 
and their stability outcome is summarized in Fig.~\ref{fig2}. As before,
Fig.~\ref{fig2} $(a_1)$ 
depicts the real part, $\rm{Re}(\omega)$, of the eigenfrequncies $\omega$ as a function of $g_{22}$,
while Fig.~\ref{fig2} $(a_2)$ shows
the corresponding imaginary part, $\rm{Im}(\omega)$.
In the real part of the spectrum the absence of the lower-lying anomalous mode is verified.
Recall that the existence of this mode in the spectrum of the antisymmetric branch 
signalled the presence of the asymmetric branch of solutions [see Fig.~\ref{fig1} $(a_1)$].
In contrast to the antisymmetric states investigated above, this family of solutions is unstable for
$g_{22}<1$ as is evident in that regime by a non-zero imaginary 
eigenfrequency illustrated in Fig.~\ref{fig2} $(a_2)$. 
On the other hand, the state remains spectrally stable for $g_{22}\geqslant 1$, 
and the formerly unstable mode, now becomes an anomalous one
with a real eigenfrequency. It is important to note here, that the equilibrium distance is found to be larger for the 
asymmetric states when compared with the antisymmetric ones [compare e.g. Fig.~\ref{fig2} $(b_2)$ with Fig.~\ref{fig1} 
$(b_2)$]. 
The equilibrium distance also becomes larger for the asymmetric states upon increasing $g_{22}$, as can be deduced by more 
closely inspecting e.g. Figs.~\ref{fig2} $(b_2)$ and $(b_3)$. 
As before, our BdG results are confirmed via 
the long-time evolution of the stationary asymmetric states and are illustrated in Figs.~\ref{fig2} $(b_1)-(c_3)$. 
Once more, Figs.~\ref{fig2} $(b_1)-(b_3)$ 
depict the evolution of the density of the dark soliton component, while Figs.~\ref{fig2} $(c_1)-(c_3)$ 
illustrate the propagation of the density of the bright soliton counterpart.         
As it is expected, for values $g_{22_{cr}}<1$ instability sets in almost from the beginning of the dynamics, with the solitons 
featuring attraction, more pronounced in the dark soliton counterpart shown in Fig.~\ref{fig2} $(b_1)$, 
which results in a collision event at intermediate time scales. 
However, and as anticipated for $g_{22}=g_{12}=1$ shown in Figs.~\ref{fig2} 
$(b_2)$ and $(c_2)$ solitons remain intact throughout the propagation, 
a result that holds as such even upon considering parameters beyond the integrable 
limit and on the miscible side depicted in Figs.~\ref{fig2} $(b_3)$ and $(c_3)$.
\begin{figure}[ht]
\includegraphics[scale=0.34]{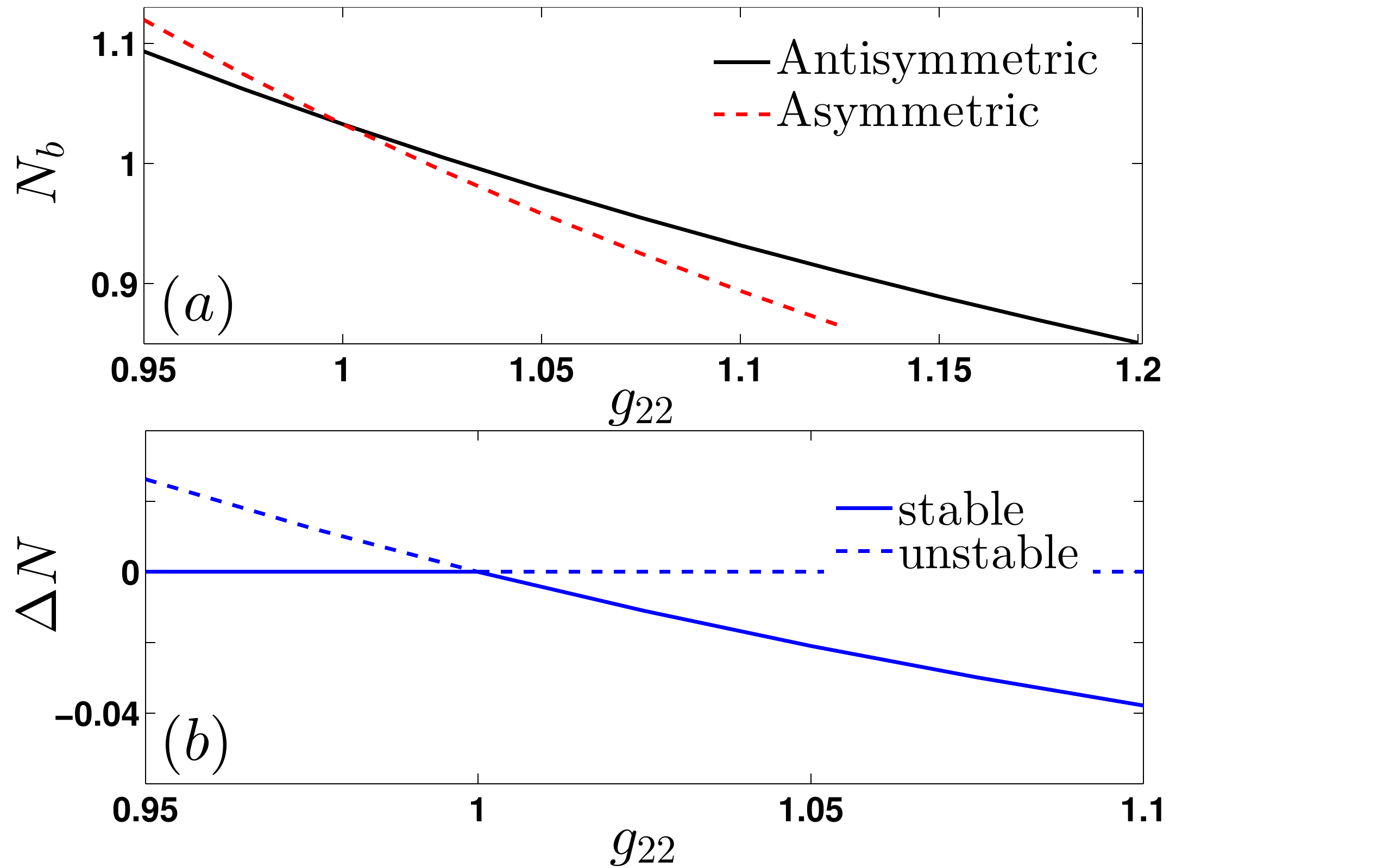}
\caption{(Color online):  
$(a)$ Number of atoms, $N_b$, of the bright soliton components
as a function of $g_{22}$ for both the antisymmetric and the asymmetric states.
$(b)$ Transcritical bifurcation diagram obtained by measuring $\Delta N$ (see text) as a function of $g_{22}$.
In all cases the stable (unstable) branches are denoted with solid (dashed) blue
lines.}
\label{fig3a}
\end{figure}

The above-observed differences between
the antisymmetric and asymmetric branches of solutions can be
further understood by inspecting the decrease in the number of atoms of the
bright component, $N_b$, as a function of the nonlinear coefficient $g_{22}$ depicted for both cases in
Fig.~\ref{fig3a} $(a)$. 
It is observed that as $g_{22}$ increases the bright norm decreases faster for the 
asymmetric states, while the two norms are exactly the same at the integrable point. 
Finally, the transcritical nature of the bifurcation diagram is 
illustrated in Fig.~\ref{fig3a} $(b)$. To obtain this bifurcation diagram we calculated the difference in the 
number of atoms in the bright part, $N_b$, for either the
asymmetric or the antisymmetric branches of solutions 
defined as $\Delta N=N_b-N^{antisym.}_b$, upon varying the nonlinear coefficient $g_{22}$. 
Notice that the stability character of the antisymmetric 
and the asymmetric states is exchanged at the integrable point verifying the
effectively transcritical nature of this bifurcation. 
It is worthwhile to comment a little on this bifurcation.
Firstly, we  point out that saddle-center
and pitchfork examples are much more common than transcritical ones
in our experience with Hamiltonian systems. In fact, the corresponding
state where $\Delta N$ possesses the opposite sign (i.e., the parity symmetric
variant of our DB-pair configuration) is also a solution. In that light,
it can be thought of as transcritical bifurcation with symmetry.
In fact, an even more crucial way in which the symmetry of the bifurcation
can be appreciated is the {\it neutrality} discussed previously at
the Manakov limit. The freedom in the variation of $\delta$ in that
context represents a one-parameter family of solutions within which
one can freely move and which represent different asymmetries in
the bright component. A by-product of this invariance is the presence
of a vanishing frequency eigenmode at the critical point of this bifurcation,
i.e., at the transition point
from stability to instability for the antisymmetric
branch or vice-versa for the asymmetric one. However, it is important
to point out that these features (neutrality, controllable asymmetry,
and associated vanishing eigenfrequency) seem to disappear once
we depart from the integrable limit, marking the absence of
additional symmetry in the latter case. 
\begin{figure*}[ht]
\centering
\includegraphics[scale=0.4]{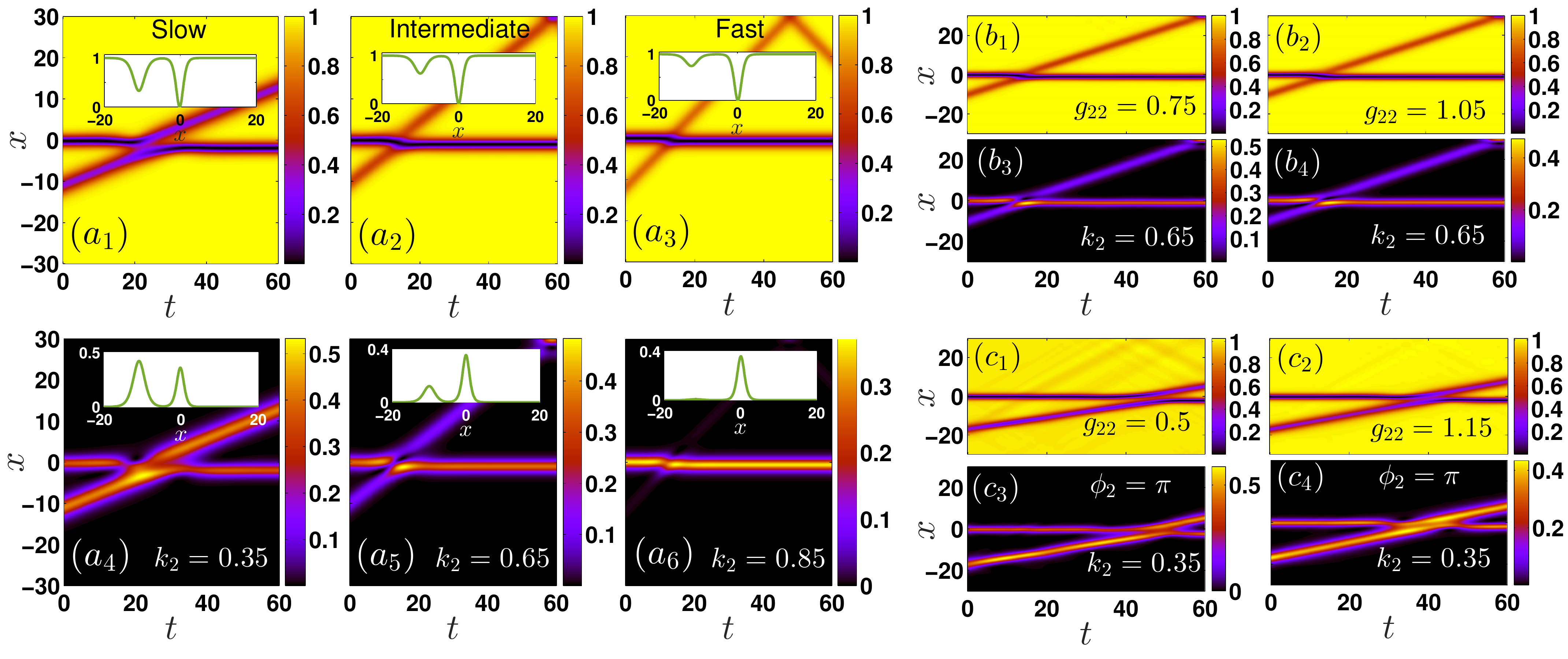}
\caption{(Color online): $(a_1)-(a_6)$  
Collisions between an initially at rest (i.e. $k_1=0$) DB pair and a moving one   
at the integrable limit. From left to right the velocity, $k_2$, of the moving DB pair is increased from $k_2=0.35$ to 
$k_2=0.85$, $\phi_1=\phi_2=\pi/4$, while the insets illustrate the initial profile of the two-DB state. 
Notice that as the velocity is increased the amplitude of the bright soliton 
decreases, leading, for $k_2=0.85$, to a state with almost only one bright soliton in the second component.
$(b_1)-(b_4)$ The same as the above but for fixed $k_2=0.65$, and for different values of the intra-species repulsion.  
$(c_1)-(c_4)$ The same as before but upon varying the phase between the soliton constituents, namely for 
$\phi_1=0$ and $\phi_2=\pi$.
In all cases the panels depict the density, $|q_1(x,t)|^2$, $\left(|q_2(x,t)|^2\right)$ of the dark (bright) soliton 
component in the top (bottom), while all quantities shown are in dimensionless units. 
Other parameters used are $\nu_1=0.8$, and $\nu_2=0.5$. }
\label{fig3}
\end{figure*}   

\section{Dark-bright soliton collisions}
In what follows we consider collisions between two-DB states at the integrable limit 
of equal inter- and intra-species interactions, i.e. $g_{12}=g_{22}=1$, as well as deviating
from it towards the miscible and the immiscible regime.
In both cases we use as an initial ansatz ($t=0$) the exact solution for such two-DB states, namely~\cite{prinari,biondini}:
\begin{widetext}
\begin{eqnarray}
q_1(x,0)&=&q_o \Bigg[1 + \frac{1}{D(x,0)} \Big[  \bar{\delta}_1 \frac{ (z_1 ^{\star})^2 } { q_o ^2 - z_1 z_2 } 
\left(  \delta_1  \frac{q_o ^2 - z_1 z_2 ^{\star} }{ z_1 (z_1 ^{\star} - z_1)} e^{-2 \nu_1 x} 
    - \delta_2 \frac{q_o ^2-|z_2|^2}{  z_2 (z_2-z_1 ^{\star}) }  e^{ ix(z_2 - z_1 ^{\star}) } \right) \nonumber \\
    &+& \frac{\bar{\delta}_2 (z_2 ^{\star})^2}{q_o^2- z_1 z_2} \left( \delta_2 \frac{q_o ^2-z_1^{\star} z_2}{z_2 (z_2^{\star}-z_2)}  
    e^{-2 \nu_2 x}                    
    - \delta_1 \frac{ q_o ^2-|z_1|^2 }{ z_1 (z_1-z_2 ^{\star}) }  e^{i x(z_1-z_2 ^{\star})} \right) \nonumber \\  
    &+& q_o ^4 |\delta_1 |^2 |\delta_2 | ^2 
       \frac{ \left( q_o ^2-|z_1|^2 \right) \left( q_o ^2-|z_2|^2 \right) |q_o ^2 - z_1^{\star} z_2|^2 |z_1-z_2|^4 
    \left( z_1^{\star} z_2 ^{\star} -  z_1 z_2 \right)}{ 16 \nu_1 ^2 \nu_2 ^2 z_1 z_2 |q_o ^2 - z_1 z_2|^2 |z_1^{\star} - z_2|^4} 
    e^{-2x \left(\nu_1 + \nu_2 \right)}  \Big] \Bigg], 
\label{twodark} \\ 
\vspace{0.5cm}
q_2(x,0)&=& -\frac{q_o}{D(x,0)} \Bigg[\frac{\bar{\delta}_1 z_1 ^{\star} }{q_o^2}  e^{-i z_1^{\star} x} 
+ \frac{\bar{\delta}_2 z_2^{\star} }{q_o^2}  e^{-iz_2 ^{\star} x} 
   + \Big[  \frac{\bar{\delta}_1 \bar{\delta}_2 z_1^{\star} z_2^{\star}
   \left(q_o^2- z_1^{\star} z_2^{\star}\right)  \left(z_1^{\star} -z_2^{\star}\right)^2 } {q_o^2 \left(q_o^2-z_1 z_2\right)}  \Big] \nonumber 
   \\ 
   &\times & \Big[ \frac{\delta_1 z_1}{\left(z_1^{\star}-z_1\right)^2 \left(z_1-z_2^{\star}\right)^2} e^{-2 \nu_1 x-i z_2^{\star}x} + 
   \frac{\delta_2 z_2}{(z_2^{\star}-z_2)^2 
   (z_2-z_1^{\star})^2}  e^{-2 \nu_2 x-i z_1^{\star} x}  \Big]\Bigg].
\label{twobright} 
\end{eqnarray}
\end{widetext}
Here, $q_1(x,0)$ [$q_2(x,0)$] is the wavefunction for the two dark [bright] soliton solution of the first [second]
component in the system of Eqs. (\ref{nls1})-(\ref{nls2}).
$q_o$ is the amplitude of the background, $z_{j}=\kappa_j+i\nu_j$   
correspond to the eigenvalues of the IST problem
where $k_j=2\kappa_j$ is the soliton's velocity,
while $\delta_{j}$ are the so-called norming
constants~\cite{prinari,biondini}. 
In all cases $j=1,2$ accounts for the first and the second component of the vector nonlinear 
Schr{\"o}dinger system. 
Additionally,
$\bar{\delta}_1=-\delta_1^{\star}( q_o^2(q_o^2-|z_1|^2) (q_o^2-z_1^{\star}z_2) )/( (z_1^{\star})^2 (q_o^2-z_1^{\star}z_2^{\star}) )$,
and $\bar{\delta}_2=-\delta_2^{\star}( q_o^2(q_o^2-|z_2|^2) (q_o^2-z_2^{\star}z_1) )/( (z_2^{\star})^2 (q_o^2-z_1^{\star}z_2^{\star}) )$,
are related to the complex conjugates of the aforementioned norming constants. 
The denominator of the above equations is 
given by
\begin{widetext}
\begin{eqnarray}
D(x,0)=1&-&\frac{\bar{\delta}_1 (z_1^{\star})^2}{q_o^2-z_1 z_2} \left(\delta_1    
 \frac{q_o^2 - z_1 z_2^{\star} } {4 \nu_1 ^2} e^{-2 \nu_1 x}
    - \delta_2 \frac{q_o^2-|z_2|^2 }{\left(z_2-z_1^{\star}\right)^2} e^{i x \left(z_2-z_1^{\star}\right)} \right) \nonumber \\
    &+& \frac{\bar{\delta}_2 (z_2^{\star})^2}{q_o^2-z_1 z_2} \left(-\delta_2 \frac{q_o^2-z_2 z_1^{\star}}{4 \nu_2^2}  e^{-2 \nu_2 x} 
    + \delta_1 \frac{q_o^2 - |z_1|^2}{\left( z_1- z_2^{\star}\right)^2} e^{i x (z_1-z_2^{\star})}  \right) \nonumber \\
    &+& q_o^4 |\delta_1 |^2 |\delta_2 |^2 
    \frac{\left(q_o^2-|z_1|^2\right) \left(q_o^2-|z_2|^2\right) |q_o^2 - z_1 z_2^{\star}|^2 |z_1-z_2|^4}{16 \nu_1^2 \nu_2 ^2 
    |q_o^2-(z_1 z_2)|^2 |z_1^{\star}-z_2|^4} e^{-2 x \left(\nu_1 + \nu_2 \right)}.
\label{denominator}
\end{eqnarray}
\end{widetext}
\begin{figure}[ht]
\centering
\includegraphics[scale=0.31]{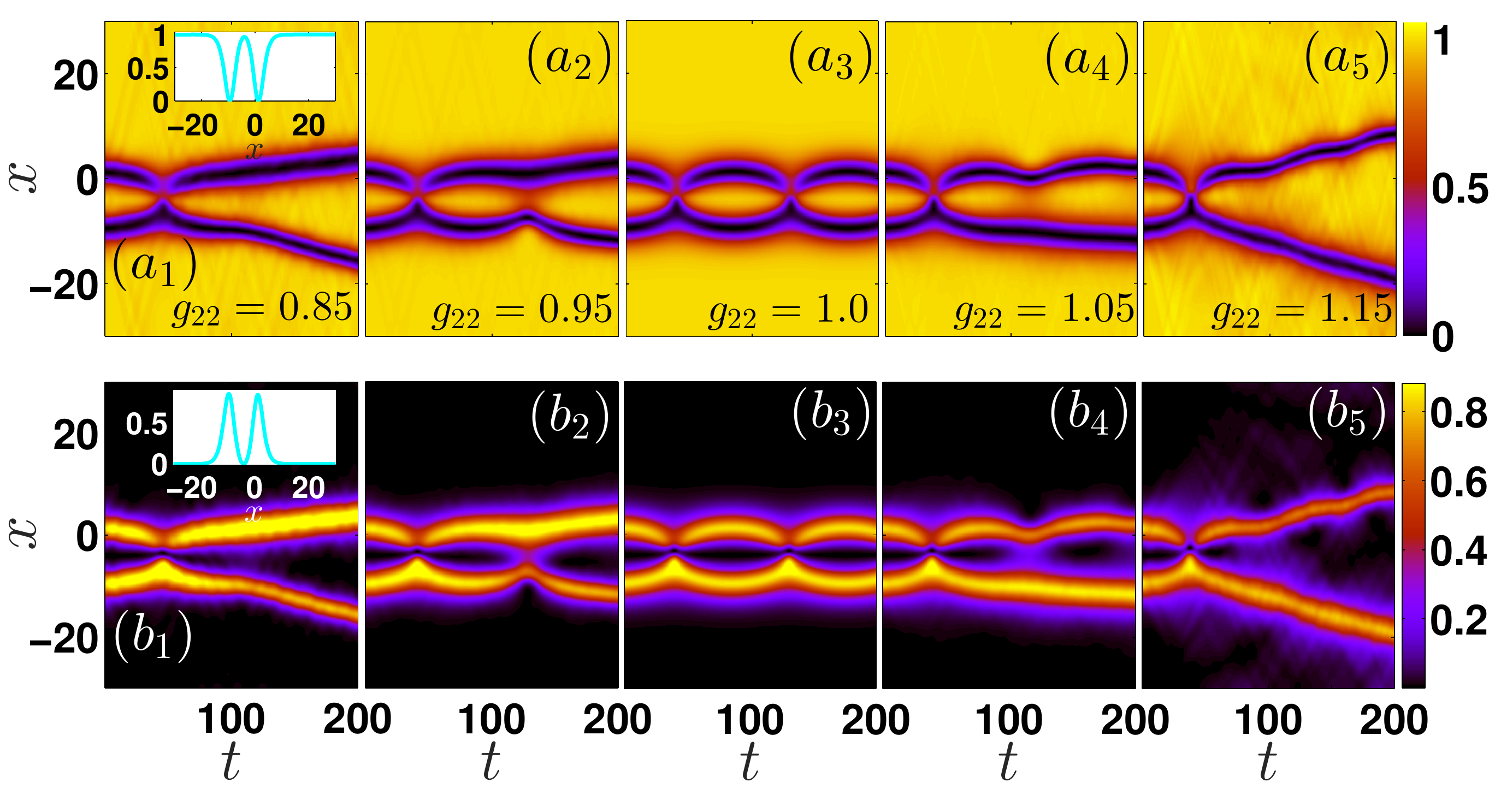}
\caption{(Color online): Collisions of DB states 
with zero initial velocities, $k_1=k_2=0$, and upon increasing 
$g_{22}$.
The corresponding density profiles for $g_{22}=0.85$ are illustrated as insets in panels $(a_1)$ and $(b_1)$.
Notice the breathing dynamics that the DB pairs undergo as the intra-species repulsion is increased.   
$(a_1)-(a_5)$ [$(b_1)-(b_5)$] Evolution of the density, $|q_1(x,t)|^2$ [$|q_2(x,t)|^2$], of the dark 
[bright] soliton component. Other parameters used are $\nu_1=1/2$, $\nu_2=1/3$, $\delta_1=1+i/5$, and $\delta_2=1-i/3$. 
All quantities shown are expressed in dimensionless units.}
\label{fig4}
\end{figure}   
In order to initialize the dynamics we first render the two-DB states of Eqs.~(\ref{twodark})-(\ref{twobright}) 
well-separated. The latter can be achieved by parametrizing the norming constants $\delta_j$ ($j=1,2$) 
as: $\delta_j=(2 \nu_j/(q_o \sqrt{q_o^2-z_j^2})) \exp(x_j+i \phi_j)$ and varying
the position offset $x_j$, and/or 
the phase $\phi_j$. Throughout this work, the amplitude of the background is fixed to $q_o=1$.  
In the case examples presented in Figs.~\ref{fig3} $(a_1)-(a_6)$, 
the DB pairs are located around $x_1\approx 0$ and $x_2\approx -10$ respectively.
Furthermore, we fix the corresponding phases 
$\phi_1=\phi_2=\pi/4$, the velocity of the first DB 
pair $k_1=0$, and we vary $k_2$ within the interval [0.35, 0.85]. It is
important to note that $k_2$ also significantly influences the asymmetry between the
bright soliton counterparts. More specifically,
for high speed solitons i.e., for $k_2>0.75$, the moving
DB has a very weak bright component gradually becoming a single dark
soliton impinging on a DB stationary wave.
We have conducted numerous collisional simulations at and below
as well as above the integrable limit (in terms of values of $g_{22}$).
Our main finding in these cases where one of the solitons possesses a
substantial speed is that generically the collisional
phenomenology remains essentially similar [see also
Figs.~\ref{fig3} $(b_1)-(b_4)$, as well as Figs.~\ref{fig3} $(c_1)-(c_4)$]
to what is predicted by the analytical expressions of the integrable
limit~\cite{prinari}. 

\begin{figure} [ht]
\centering
\includegraphics[scale=0.31]{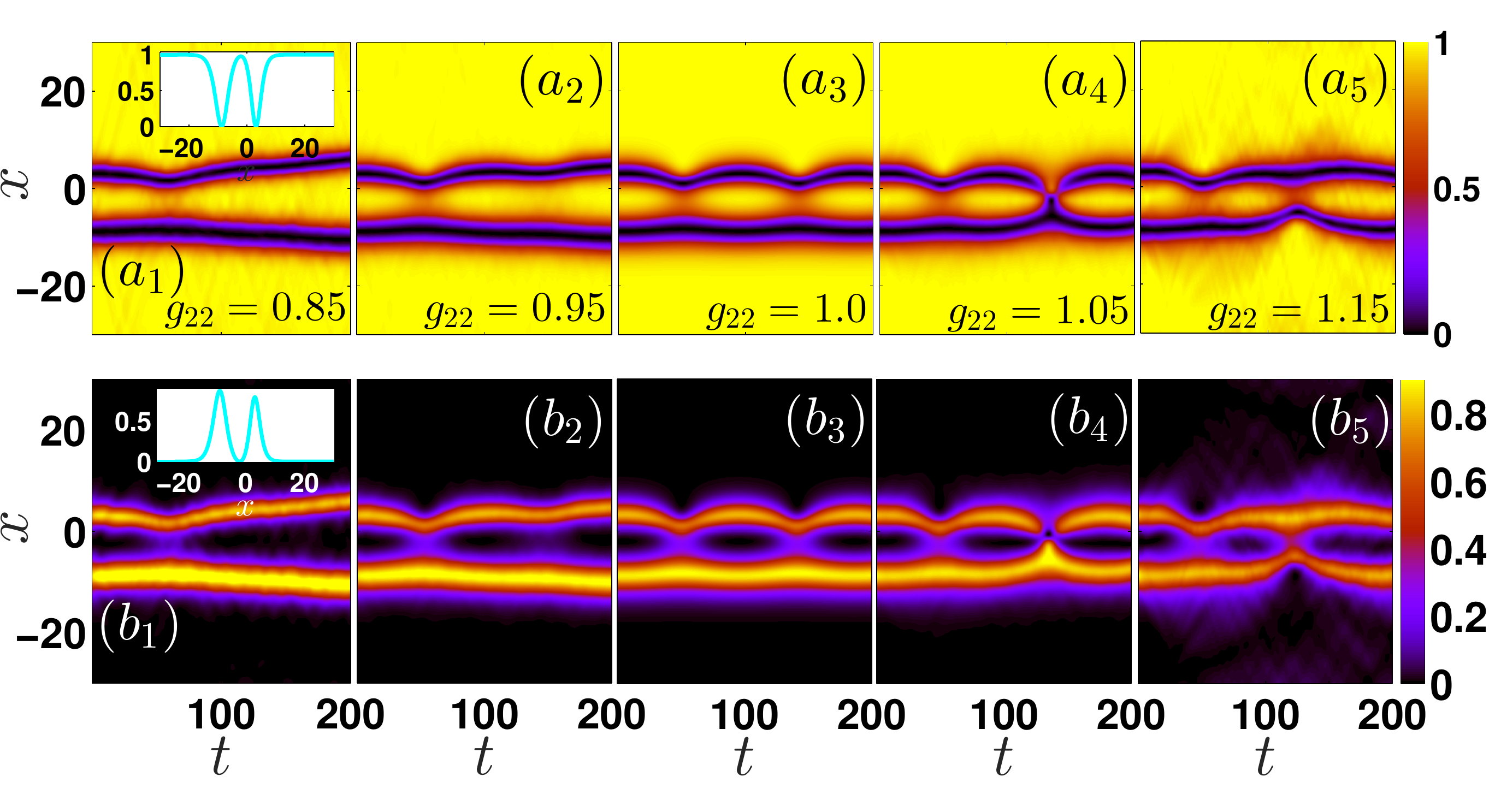}
\caption{(Color online): Same as in Fig.~\ref{fig4} but for $\delta_1=5+i/5$, 
with all quantities shown expressed in dimensionless units.}
\label{fig5}
\end{figure}

This situation is in contrast to cases where the solitons have been initialized
with vanishing speed, in which we have seen that the breaking of integrability
has a maximal impact. A characteristic example of this kind is given
in Figs.~\ref{fig4} $(a_1)-(b_5)$, where once more Figs.~\ref{fig4} $(a_1)-(a_5)$ [$(b_1)-(b_5)$]
depict the breathing dynamics of the dark [bright] soliton constituent. 
In this case involving $k_1=k_2=0$, and the choice of the norming
constants of $\delta_1=1+i/5$ and $\delta_2=1-i/3$ (for $\nu_1=1/2$
and $\nu_2=1/3$), we find that at the integrable limit
the solution forms a beating state. This ``fragile'' beating
is already seen to be significantly impacted by small deviations
from integrability of the order of $5 \%$ as it is evident in 
Figs.~\ref{fig4} $(a_2), (b_2)$, and $(a_4),(b_4)$, which refer to deviations 
towards the miscible and the immiscible regime 
respectively. However, the phenomenology
is dramatically affected for deviations of the order of $15 \%$ or
more, whereby the former beating state gives way, upon already
the first collision of the DB pair, to an indefinite separation
between the two DBs. Remarkably, this deviation takes place
both in the miscible, Figs.~\ref{fig4} $(a_1), (b_1)$, and in the immiscible regime,
Figs.~\ref{fig4} $(a_5), (b_5)$.
Under different values of the norming constants, the departure from
the breathing state may be ``decelerated''.
A case example of this kind is depicted in
Fig.~\ref{fig5}. In particular in this realization all parameters used
are the same as in the aforementioned collisional scenario 
except for $\delta_1=5 + i/5$. 
Notice that for this choice of parameters the aforementioned deceleration against repulsion is 
more pronounced within the immiscible regime of interactions [compare e.g. 
panels $(a_5)$ and $(b_5)$ here, with Figs.~\ref{fig4} $(a_5)$ 
and $(b_5)$].
Furthermore, this change in the norming 
constant affects the bright soliton characteristics resulting in particular 
in slightly mass imbalanced bright soliton counterparts, illustrated as insets in Figs.~\ref{fig5} $(a_1)$ and $(b_1)$, 
when compared to the spatial profiles of the solitons shown as insets in Figs.~\ref{fig4} $(a_1)$ and $(b_1)$.   
On the other hand, when initializing 
the dynamics considering a genuinely stationary configuration, the results of
the stability analysis of the previous section are confirmed.
Namely, for a genuinely antisymmetric bright configuration, stability
persists on the side of $g_{22}<1$, while a splitting (symmetry breaking)
instability into a DB and a dark soliton arises for $g_{22}>1$, with the
stability intervals being reversed for asymmetric initial conditions.

\section{Conclusions and future challenges}

In the present work we have investigated 
the stability and dynamics of matter-wave DB solitons
in homogeneous binary BECs. We have
done so by taking advantage in a systematic fashion of the understanding
and knowledge imparted by the inverse scattering transform and the
Hirota method within the integrable Manakov limit. In that limit,
both antisymmetric and asymmetric DB pair waveforms are identified,
but also solutions involving DB pairs interacting with different speeds
can be explored via cumbersome, yet explicit analytically available
formulae. We have benchmarked the results of our numerical simulations
against these expressions and subsequently extended our analysis
past the integrable limit to identify the nature of the deviations
from the corresponding results. This has given rise to an array of
interesting findings. In particular, in the case of stationary solutions
we have identified an intriguing transcritical bifurcation with symmetry.
Antisymmetric and asymmetric solutions in the bright soliton component
of the DB pairs have been found, respectively to be, stable (unstable)
for $g_{22}<1$ ($g_{22}>1$), exchanging their stability in the degenerate
(invariant under symmetry breaking) case of the integrable limit.
Moreover, as regards collisions, we have seen that those bearing
significant kinetic energy were essentially unaffected by the breaking
of integrability. On the other hand, the more delicate beating
(or even stationary) states were drastically affected by the breaking
of integrability, typically leading to the fission of the elements
within the pair, possibly accompanied by a symmetry breaking between
the bright components.

These findings suggest a multitude of interesting directions for future studies.
A straight forward one would be to consider such multicomponent interactions in the
presence of quantum fluctuations~\cite{lgs}. In such a setting it has recently been shown 
that DB states decay into daughter DB ones, so it would be particularly 
interesting to explore how the collisional dynamics of the above beating states
is altered by taking into account beyond mean-field effects.
Yet another interesting aspect would be to extend our current considerations involving 
a higher number of species. In this spinor setting, solutions in the form of dark-dark-bright
and dark-bright-bright solitons have been theoretically obtained~\cite{BG}, and also very recently experimentally 
observed~\cite{PGK_Engels}. Thus a study of their static and dynamical properties
will enhance our understanding of these soliton complexes.
Furthermore, one could also explore multicomponent interactions as the ones considered 
herein but in higher dimensions. 
As is well-known there are no direct analogues
of the Manakov model that are known to be integrable at present. However,
it is nevertheless of interest to explore interactions of vortex-bright
solitons in two dimensions~\cite{pola} and of configurations
such as vortex line-bright solitons or vortex-ring-bright solitons
in three dimensions~\cite{wenlong}. These possibilities are presently
under consideration and will be reported in future publications.

\section*{Acknowledgements} 

P.G.K. gratefully acknowledges the support of NSF-PHY-1602994, and the
Alexander von Humboldt Foundation. B.P. gratefully acknowledges
the support of NSF-DMS-1614601, and G.B. gratefully acknowledges
the support of NSF-DMS-1614623, NSF-DMS-1615524.   
G.C.K and P.S. gratefully acknowledge fruitful discussions with J. Stockhofe. 




\begin{thebibliography}{99}

\bibitem{gpes1}  L. P. Pitaevskii, and S. Stringari, 
{\it Bose-Einstein Condensation},
Oxford University Press, (Oxford, 2003).

\bibitem{gpes2} P. G. Kevrekidis, D. J. Frantzeskakis, and R. Carretero-Gonz{\'a}lez, 
{\it Emergent Nonlinear Phenomena in Bose-Einstein Condensates},
Springer-Verlag (Berlin, 2008).

\bibitem{frantzkevre} P. G. Kevrekidis, and D. J. Frantzeskakis, Reviews in Physics {\bf 1}, 140 (2016).


\bibitem{hamburg} 
C. Becker, S. Stellmer, P. Soltan-Panahi, S. D{\"o}rscher, M. Baumert, E.-M.
Richter, J. Kronj\"{a}ger, K. Bongs, and K. Sengstock, 
Nat. Phys. \textbf{4}, 496 (2008).

\bibitem{yan} D. Yan, J. J. Chang, C. Hamner, P. G. Kevrekidis, P. Engels, V. Achilleos, D. J. Frantzeskakis, 
R. Carretero-Gonz{\'a}lez, P. Schmelcher, 
Phys. Rev. A  {\bf 84}, 053630 (2011).

\bibitem{ba} T. Busch, and J. R. Anglin, 
Phys. Rev. Lett. {\bf 87}, 010401 (2001).

\bibitem{hamner} C. J. Hamner, J. J. Chang, P. Engels, and M. A. Hoefer, 
Phys. Rev. Lett. {\bf 106}, 065302 (2011).

\bibitem{middelkamp} S. Middelkamp, J. J. Chang, C. Hamner, R. Carretero-Gonz{\'a}lez, P. G. Kevrekidis, V. Achilleos,
D. J. Frantzeskakis, P. Schmelcher, and P. Engels,  
Phys. Lett. A  {\bf 375}, 642 (2011).

\bibitem{alvarez} A. {\'A}lvarez, J. Cuevas, F. R. Romero, C. Hamner, J. J. Chang, P. Engels, P. G. Kevrekidis, and D. J. 
Frantzeskakis, 
J. Phys. B  {\bf 46}, 065302 (2013).


\bibitem{nls}  M. J. Ablowitz, B. Prinari, and A. D. Trubatch,
{\it Discrete and Continuous Nonlinear Schr{\"o}dinger Systems}, Cambridge University Press
(Cambridge, 2004).

\bibitem{siambook} P.~G.~Kevrekidis,
D.~J.~Frantzeskakis, and R.~Carretero-Gonz{\'a}lez,
{\it The Defocusing Nonlinear Schr{\"o}dinger Equation},
SIAM (Philadelphia, 2015).


\bibitem{manakov} S. V. Manakov, Zh. Eksp. Teor. Fiz. {\bf 65}, 505 (1973)
[Sov. Phys. JETP {\bf 38}, 248 (1974)].


\bibitem{prinari}  D. Garrett,  T. Klotz, B. Prinari, and F. Vitale,
Applic. Anal. {\bf 92}, 379 (2013).

\bibitem{biondini}  B. Prinari, F. Vitale, and G. Biondini,
J. Math. Phys. {\bf 56}, 071505 (2015).

\bibitem{shepkiv} 
A. P. Sheppard, and Yu. S. Kivshar, 
Phys.\ Rev.\ E \textbf{55}, 4773 (1997).

\bibitem{vdbysk1} 
V. V. Afanasjev, Yu. S. Kivshar, V. V. Konotop, and V. N. Serkin, 
Opt. Lett. \textbf{14}, 805 (1989).

\bibitem{ralak} 
R. Radhakrishnan, and M. Lakshmanan,
J.\ Phys.\ A: Math.\ Gen. \textbf{28}, 2683 (1995).

\bibitem{parkshin} 
Q. -H. Park and H. J. Shin, 
Phys.\ Rev.\ E \textbf{61}, 3093 (2000).

\bibitem{rajendran} S. Rajendran, P. Muruganandam, and M. Lakshmanan,
  J. Phys. B {\bf 42}, 145307  (2009).

\bibitem{inouye}  S. Inouye, M. R. Andrews, J. Stenger, H.-J. Miesner D. M. Stamper-Kurn, and W. Ketterle, 
Nature (London) {\bf 392}, 151 (1998).

\bibitem{roberts} J. L. Roberts, N. R. Claussen, J. P. Burke, Jr., C. H. Greene, E. A. Cornell, 
and C. E. Wieman, 
Phys. Rev. Lett. {\bf 81}, 5109 (1998). 


\bibitem{donley} E. A. Donley, N. R. Claussen, S. L. Cornish, J. L. Roberts, E. A. Cornell, and C. E. Wieman, 
Nature (London) {\bf 412}, 295 (2001).


\bibitem{thalhammer} G. Thalhammer, G. Barontini, L. de Sarlo, J. Catani, F. Minardi, and M. Inguscio, 
Phys. Rev. Lett. {\bf 100}, 210402 (2008).

\bibitem{chin} C. Chin, R. Grimm, P. Julienne, and E. Tiesinga, 
Rev. Mod. Phys. {\bf 82}, 1225 (2010).

\bibitem{ef} D. Yan, F. Tsitoura, P. G. Kevrekidis, and D. J. Frantzeskakis,
Phys. Rev. A {\bf 91}, 023619 (2015).

\bibitem{et} E. T. Karamatskos, J. Stockhofe, P. G. Kevrekidis, and P. Schmelcher, 
Phys. Rev. A {\bf 91}, 043637 (2015).

\bibitem{ljpp} G. C. Katsimiga, J. Stockhofe, P. G. Kevrekidis, and P. Schmelcher, 
Phys. Rev. A {\bf 95}, 013621 (2017).

\bibitem{carr} O. Majed, D. Alotaibi, and L. D. Carr, 
Phys. Rev. A {\bf 96}, 013601 (2017).

\bibitem{ljpp1} G. C. Katsimiga, J. Stockhofe, P. G. Kevrekidis, and P. Schmelcher, 
Appl. Sci. {\bf 7}, 388 (2017).


\bibitem{aochui} P. Ao, and S. T. Chui,
Phys. Rev. A {\bf 58}, 4836 (1998).

\bibitem{krein} D. V. Skryabin,
Phys. Rev. A {\bf 63}, 013602 (2000).

\bibitem{lgs}   G. C. Katsimiga, G. M. Koutentakis,  S. I. Mistakidis,
P. G. Kevrekidis, and P. Schmelcher, New J. Phys.
{\bf 19} 073004 (2017).

\bibitem{BG}  G. Biondini, D. K. Kraus, and B. Prinari,
Commun. Math. Phys. {\bf 348}, 475 (2016).
 
\bibitem{PGK_Engels} T. M. Bersano, V. Gokhroo, M. A. Khamehchi, J. D' Ambroise, D. J. Frantzeskakis, P. Engels, 
and P. G. Kevrekidis, {\bf arXiv:1705.08130} (2017).
 
 
\bibitem{pola} M. Pola, J. Stockhofe, P. Schmelcher, and P. G. Kevrekidis,
  Phys. Rev. A {\bf 86}, 053601 (2012).

\bibitem{wenlong} E. G. Charalampidis, W. Wang, P. G. Kevrekidis, D. J. Frantzeskakis, and J. Cuevas-Maraver,
Phys. Rev. A {\bf 93}, 063623 (2016). 
  
\end{thebibliography}
\end{document}